\documentclass[11pt,english]{article}
\usepackage[T1]{fontenc}
\usepackage[latin9]{inputenc}
\usepackage{geometry}
\usepackage{babel}
\usepackage{float}
\usepackage{amsmath}
\usepackage{amssymb}
\usepackage{graphicx}
\usepackage{esint}
\usepackage[unicode=true,pdfusetitle,
 bookmarks=true,bookmarksnumbered=false,bookmarksopen=false,
 breaklinks=false,pdfborder={0 0 1},backref=false,colorlinks=false]
 {hyperref}

\makeatletter
\newcommand{\lyxaddress}[1]{
	\par {\raggedright #1
	\vspace{1.4em}
	\noindent\par}
}

\makeatother

\begin{document}
\title{Periodically driven perturbed CFTs: the sine-Gordon model}
\author{Zoltán Bajnok$^{1}$, Robin Oberfrank$^{1,2}$}
\maketitle

\lyxaddress{\begin{center}
$^{1}$\emph{Wigner Research Centre for Physics}\\
\emph{Konkoly-Thege Miklós u. 29-33, 1121 Budapest, Hungary}\\
$^{2}$\emph{Roland Eötvös University}\\
\emph{Pázmány s. 1/A, 1117 Budapest, Hungary}\\
\emph{}\\
\par\end{center}}
\begin{abstract}
We analyze a version of the sine-Gordon model in which the strength
of the cosine potential has a periodic dependence on time. This model
can be considered as the continuum limit of the many body generalization
of the Kapitza pendulum. Based on the perturbed CFT point of view,
we develop a truncated conformal space approach (TCSA) to investigate
the Floquet quasienergy spectrum. We focus on the effective behaviour
for large driving frequencies, which we also derive exactly. Depending
on the driving protocol, we can recover the original sine-Gordon model
or its two-frequency version. The rich structure of the two-frequency
model implies that the time-periodic drive can break integrability,
can lead to new states in the spectrum or can result in a phase transition.
Our method is applicable for any periodically driven perturbed conformal
field theories.
\end{abstract}
\newpage{}

\tableofcontents{}

\section{Introduction}

The equilibrium behaviour of isolated statistical physical systems
are successfully described and understood \cite{Mussardo:2020rxh}.
Recently, the focus of the investigations has been shifted to the
far from equilibrium domain, partly due to the advancement of cold
atom experiments. By introducing various protocols, the system can
be driven away from the equilibrium and the main problem is to understand
its long time behaviour. Typically, we let the originally closed quantum
system to interact with its environment. This can be a sudden change
in the boundary conditions or in the parameters of the theory and
often we close the system again after this quench and we investigate
the relaxation towards equilibrium \cite{Quench}.

Alternatively, we can subject the system to a periodic driving force
and investigate its time evolution. In many cases, the driven interacting
system becomes ergodic and visits the whole phase space in the classical
case or heats to infinite temperature in the quantum case. However,
there is a large class of models when this does not happen and the
system remains stable. The periodic drive can lead to universal high
frequency behaviour, which leads to dynamical stabilization and can
be used in Floquet engineering, a topic very intensively investigated
recently \cite{Bukov_2015}.

A particularly interesting case is when the originally unstable fix
points become stable. A prototypical example is the Kapitza pendulum
\cite{Kapitza}, a rigid pendulum in which the pivot point is moved
harmonically in the vertical direction. For large enough driving frequencies
the upper unstable equilibrium point can be dynamically stabilized
by this periodic drive. The field theoretical limit of the many-body
generalization of the Kapitza pendulum is the periodically driven
sine-Gordon model \cite{Citro_2015}, which was analyzed by various
approximate methods and the dynamical stability was indicated. The
authors also suggested an experimental realization in an ultracold
atom experiment, in which the atoms are trapped into two parallel
one dimensional lines by a transversal confinement potential generated
through standing laser waves. By modulating the amplitude of the transverse
field a time-dependent tunneling coupling between the two parallel
tubes can be induced realizing the drive needed for the sine-Gordon
model.

In the present paper, we would like to go beyond the methods and the
focus of \cite{Citro_2015} and would like to analyze the quasienergy
spectrum of the driven sine-Gordon theory. This will be done by exploiting
that the sine-Gordon theory can be considered as a perturbed conformal
field theory (CFT). This allows us to use analytical and very successful
numerical methods to investigate the system. The truncated conformal
space approach (TCSA) \cite{Yurov:1989yu} proved to be a useful tool
to analyze various observables in perturbed conformal field theories
and our aim is to combine it with the usual numerical method of the
periodically driven systems to calculate the spectrum of the Floquet
Hamiltonian. We are going to identify a protocol when the infinite
frequency effective theory becomes the two-frequency sine-Gordon model
\cite{Delfino:1997ya,Bajnok:2000ar} allowing us to see the dynamical
stabilization of unstable fix points and reach very nontrivial limiting
theories. Our approach paves the way to investigate periodically driven
perturbed conformal field theories with the same methods. In the formulation
of our approach we take a pedagogical path and introduce all concepts
in simplified circumstances.

The paper is organized as follows: we start in section 2 by recalling
the dynamical stabilization and separation of time scales in the example
of the Kapitza pendulum, which is basically the zero mode of the driven
sine-Gordon theory. We also emphasize the possibility of introducing
different driving protocols, depending on how we scale the amplitude
of the drive with the frequency. By coupling many Kapitza pendula
and taking their continuum limit, we arrive at the driven sine-Gordon
theory, whose stability we analyze in the small amplitude limit, where
we can see that stability can be ensured only in a finite volume or
with a momentum cutoff. We then turn to the quantum theory in section
3. Floquet theory, the quantum approach to periodically driven systems
is introduced on the example of the quantum Kapitza pendulum, together
with the standard numerical method based on Fourier transformation
to calculate the quasienergy spectrum and an analytical method to
derive the large frequency expansion \cite{PhysRevA.68.013820}. Having
introduced the sine-Gordon theory as a perturbed CFT together with
its TCSA method \cite{Feverati:1998va}, we specify the previous findings
to deal with the periodical drive. We also perform analytical calculations
and reveal the importance of scaling the amplitude of the drive. The
numerical investigations and their interpretations are summarized
in section 4. The analytical calculation for the large frequency effective
behaviour is extended for generic perturbed conformal field theories
in Section 5. Finally, we conclude in section 6. Technical details
are relegated to Appendices.

\section{Classical considerations}

In this section, we review the classical theory and introduce the
separation of scales as well as the large frequency expansion.

\subsection{Kapitza pendulum}

\begin{figure}[h]
\begin{centering}
\includegraphics[width=10cm]{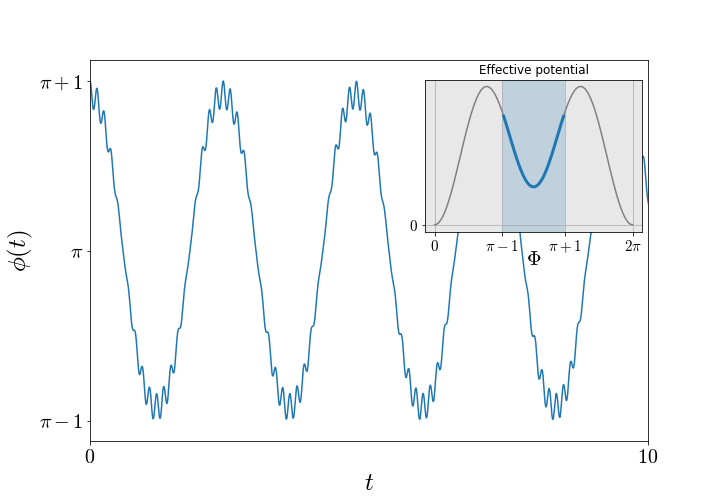}
\par\end{centering}
\caption{Periodic motion around the upper equilibrium point obtained by solving
the equation of motions with the periodic drive, and the corresponding
effective potential.}

\label{fig:pi}
\end{figure}

The Kapitza pendulum is a rigid pendulum in which the pivot point
is moved harmonically in the vertical direction \cite{Kapitza}. The
Hamiltonian has the form 
\begin{equation}
H=\frac{p_{\phi}^{2}}{2}+c(t)(1-\cos\phi)\quad;\qquad c(t)=c_{0}+c_{1}\cos\omega t
\end{equation}
where $\phi$ is the $2\pi$ periodic angle variable and the driving
is controlled by $c_{1}$. Without the drive, the system has two equilibrium
points: $\phi=0$ is stable, while $\phi=\pi$ is unstable. For large
enough driving frequencies, however, the unstable fix point becomes
stable. This counterintuitive phenomenon can be understood in the
effective description, in which the motion is separated into an average
slow and a periodic fast motion
\begin{equation}
\phi(t)=\Phi(t)+\xi(t)\quad;\qquad\xi(t+T)=\xi(t)
\end{equation}
where $T=\frac{2\pi}{\omega}$ and $\int_{t-T/2}^{t+T/2}\phi(t)dt=\Phi(t)$.
There is a systematic expansion in $\omega^{-2}$, see \cite{PhysRevA.68.013820}
for details, which at the leading order gives
\begin{equation}
\xi=\frac{c_{1}}{\omega^{2}}\cos\omega t\sin\Phi\quad;\qquad\ddot{\Phi}=-\frac{d}{d\Phi}\left(c_{0}(1-\cos\Phi)+\frac{c_{1}^{2}}{4\omega^{2}}\sin^{2}\Phi\right)\equiv-\frac{dU_{\mathrm{eff}}(\Phi)}{d\Phi}\label{eq:Ueff}
\end{equation}
 Thus, there is a small-amplitude, large-frequency motion with vanishing
time average and a slow motion in an effective potential. The new
term in the effective potential makes the upper equilibrium point
stable as demonstrated on Figure \ref{fig:pi}.

\begin{figure}[h]
\begin{centering}
\includegraphics[width=7cm]{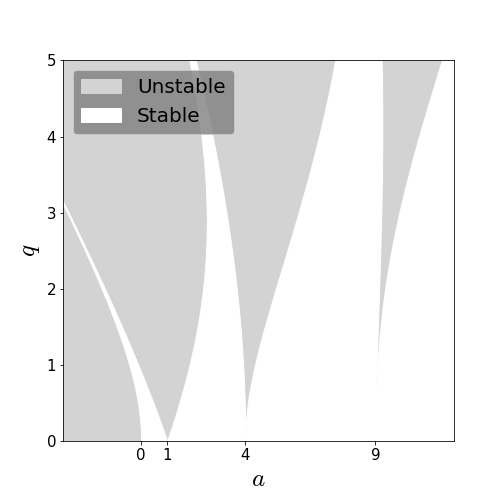}
\par\end{centering}
\caption{Stability diagram of the Mathieu equation. Stable regions correspond
to bounded oscillations, while unstable ones mean exponentially growing
solutions.}

\centering{}\label{fig:mathieu}
\end{figure}

Depending on how the amplitude of the drive scales with $\omega$,
we can see different behaviours:
\begin{itemize}
\item For $\omega$-independent $c_{1}=\lambda$, the drive averages out
in the large $\omega$ limit, and the effective motion is the same
as the one without the drive. It can be understood physically as the
drive is oscillating so fast so that the system with a finite inertia
cannot follow it.
\item This is drastically changed if the drive scales with $\omega$: $c_{1}=\lambda\omega$.
In this case, the fast oscillation vanishes in the large $\omega$
limit, and the motion is basically as it would happen in the effective,
$\omega$-independent potential $U_{\mathrm{eff}}(\Phi)=c_{0}(1-\cos\Phi)+\frac{\lambda^{2}}{4}\sin^{2}\Phi$.
For $\frac{\lambda^{2}}{2}>c_{0}$, both equilibria are stable.
\item In the case of the Kapitza pendulum the drive is proportional to $\omega^{2}$:
$c_{1}=\lambda\omega^{2}$. The small fluctuations are $\omega$-independent,
but the effective potential does not have a finite $\omega\to\infty$
limit. In this case, $\omega$ is kept finite and the system has a
rich stability diagram.
\end{itemize}
The stability of the two equilibria can be understood in the small-angle
limit, $\phi\sim\epsilon,\pi+\epsilon$, i.e $\sin\phi\sim\pm\epsilon$
when the equation of motion can be mapped to the Mathieu equation
\begin{equation}
y''(x)+(a-2q\cos2x)y(x)=0\quad;\quad(a,q,x)\leftrightarrow\left(\pm\frac{4c_{0}}{\omega^{2}},\frac{2c_{1}}{\omega^{2}},\frac{\omega t}{2}\right)
\end{equation}
with a well-known stability diagram (see Figure \ref{fig:mathieu}).
The upper equilibrium appears for $a<0$, while the lower one for
$a>0$. Clearly, for the Kapitza pendulum $c_{1}=\lambda\omega^{2}$,
the upper fix point becomes stable above a $\lambda$-dependent critical
frequency.

In the following sections, we focus on the case when the effective
potential has a finite non-trivial $\omega\to\infty$ limit, and investigate
the continuum limit of the many body generalization of this model.

\subsection{Many body generalization: the sine-Gordon model}

If we take many pendula on a line, couple them with torsion springs
and take the continuum limit we obtain the sine-Gordon field theory
with infinite degrees of freedom, see eg. \cite{Samaj:2013yva}. Doing
the same with many Kapitza pendula leads to the driven sine-Gordon
model \cite{Citro_2015}:

\begin{equation}
\partial_{t}^{2}\phi(x,t)-\partial_{x}^{2}\phi(x,t)+(c_{0}+c_{1}\cos\omega t)\sin\phi(x,t)=0
\end{equation}
The stability of the system around the $\phi(x,t)=0,\pi$ configurations
can be analyzed by approximating $\sin\phi\sim\pm\phi$ and decoupling
the modes in Fourier space. The resulting equation of motion for the
$k$th mode $\phi(x,t)\propto\phi_{k}(t)e^{ikx}$ can again be mapped
to the Mathieu equation 
\begin{equation}
\partial_{t}^{2}\phi_{k}\pm(c_{0}\pm k^{2}+c_{1}\cos\omega t)\phi_{k}=0\quad;\qquad(a,q)\leftrightarrow\left(4\frac{\pm c_{0}+k^{2}}{\omega^{2}},\frac{2c_{1}}{\omega^{2}}\right)
\end{equation}

\begin{figure}[h]
\begin{centering}
\includegraphics[width=6cm]{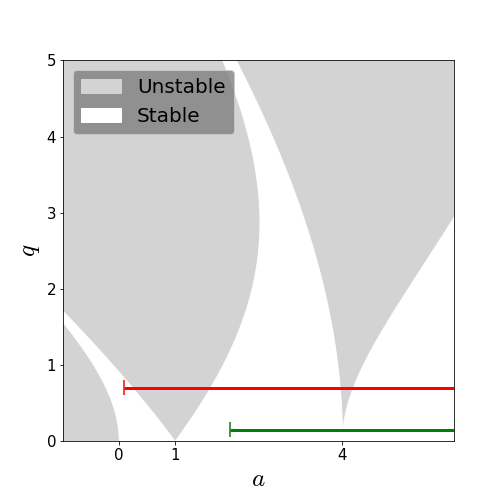}~~~~~~\includegraphics[width=6cm]{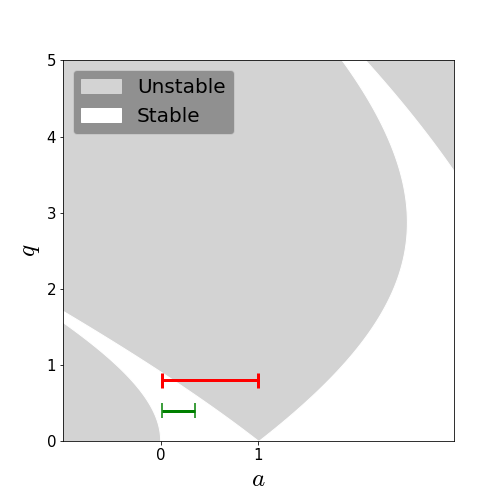}~~~~~~
\par\end{centering}
\caption{Small fluctuations of the periodically driven sine-Gordon equation
are mapped to the stability diagram of the Mathieu equation for various
parameters. Red lines correspond to evolving field configurations
with resonant modes, green ones have stable evolution. Left figure
shows solutions localized to the lower fix point, where the momentum
cutoff is not necessary, while on the right, we see solutions localized
to the upper fix point, where the cutoff has to be used to avoid the
resonant modes.}

\label{fig:dsGinst}
\end{figure}

Since in the field theory, we have a continuum of modes $k\in\mathbb{R}$,
whatever $c_{0}$ and $c_{1}$ we choose we always cross the instability
regions, see Figure \ref{fig:dsGinst}. The reason is that in our
small-field decoupled harmonic oscillator limit, we always have a
mode, which resonates with the driving frequency leading to parametric
resonance. To avoid this, we could put the system in a finite volume
$L$, and then the possible $k$ values will be quantized: $k_{n}=\frac{n2\pi}{L}$,
and we might avoid the discrete points that lie within the unstable
regions. In the large volume limit, however, we should always face
some instabilities. We could also introduce a momentum cutoff $k_{\mathrm{max}}$,
and choose a large enough driving frequency, such that the corresponding
allowed $(a,q)$ values always lie in the stability region. This is
actually the typical case as in many applications, the sine-Gordon
model is an approximation, and the real physical system has a built
in ultra-violet cutoff.

\begin{figure}[h]
\begin{centering}
\includegraphics[width=16cm]{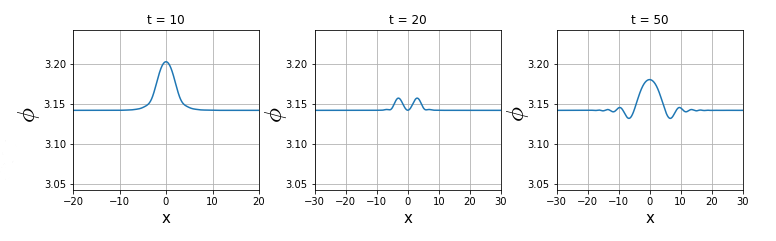}\caption{Time evolution of a breather type configuration at the upper fix point
with parameters: $c_{1}=\lambda\omega^{2}$, $\lambda=0.1$, $\omega=100$,
$c_{0}=1$. This configuration does not decay, but slowly radiates
due to non-integrable effects \cite{Fodor:2008es}.}
\par\end{centering}
\label{fig:stableBreather}
\end{figure}

With this trick, we can even make the upper fix point, $\phi=\pi$
stable. This configuration then can serve as another vacuum and excitation
such as the breather can live over them. We demonstrate this by explicitly
solving the equation of motion (see Figure \ref{fig:stableBreather}).
In our approach, we used the Chebyshev spectral method to calculate
the time evolution.

It is a conceptually interesting question whether the stability survives
in the quantum theory or not. Clearly at the quantum level, the excitations
are quantized and they cannot be arbitrarily small, thus the small
$\phi$ expansion is not adequate. Indeed, in the sine-Gordon theory,
the small fluctuations correspond to breather excitations, which are
the elementary quanta of the field with finite masses. Depending on
the coupling, the masses vary and the breathers can even leave the
spectrum, such that only solitons remain. This makes the quantum theory
stable as we will see. In the quantum analyzis, we will introduce
both a finite volume and a momentum cutoff, and will focus on the
large-frequency limit, varying both the volume and the cutoff, too.

\section{Quantum models}

In this section, we introduce the Floquet theory of periodically driven
systems together with their large frequency expansion and develop
a numerical method to investigate the spectrum of the quasienergies.
We start with the quantum version of the Kapitza pendulum. We then
present the sine-Gordon theory, where we exploit its perturbed CFT
formulation.

\subsection{Floquet theory and the Kapitza pendulum}

In the quantum theory, we investigate the time evolution of the system
with an explicit harmonic-type time-dependent Hamiltonian

\begin{equation}
i\frac{\partial\Psi}{\partial t}=H\Psi\quad;\qquad H=H_{0}+V_{0}+\cos\omega t\,V_{1}\label{eq:Hdriven}
\end{equation}
i.e. $V_{0},V_{1}$ will not depend explicitly on time. In the case
of the Kapitza pendulum, they take the form
\begin{equation}
H_{0}=\frac{p_{\phi}^{2}}{2}\quad;\quad V_{0}=c_{0}(1-\cos\phi)\quad;\quad V_{1}=c_{1}\cos\phi
\end{equation}
The separation of scales has a quantum analogue \cite{PhysRevA.68.013820},
which is phrased in Floquet theorem: for periodic Hamiltonians $H(t)=H(t+T)$
the solution of the Schrödinger equation can be expanded in a basis
$\psi_{\mathcal{E}}$ evolving as 
\begin{equation}
\psi_{\mathcal{E}}(\phi,t)=e^{-i\mathcal{E}t}u_{\mathcal{E}}(\phi,t)
\end{equation}
where $u_{\mathcal{E}}(\phi,t+T)=u_{\mathcal{E}}(\phi,t)$. The quasienergy
$\mathcal{E}$ governs the slow motion in an effective potential,
while the periodic $u_{\mathcal{E}}$ is the analogue of the fast
modulation. In order to find the effective Hamiltonian, which is unitary
equivalent to the operator that generates the evolution with one period
of time, one can make a periodic gauge transformation \cite{PhysRevA.68.013820}
of the form $u_{\mathcal{E}}(\phi,t)=e^{-iF(t)}v_{\mathcal{E}}(\phi)$
leading to
\begin{equation}
H_{\mathrm{eff}}=e^{iF}He^{-iF}-i(\partial_{t}e^{iF})e^{-iF}
\end{equation}
where $F$ and $H_{\mathrm{eff}}$ can be calculated simultaneously
order by order in $\omega^{-1}$, see appendix \ref{sec:Large-frequency-expansion},
and at the leading and non-vanishing subleading order we get 
\begin{equation}
H_{\mathrm{eff}}=H_{0}+V_{0}+\frac{1}{4\omega^{2}}[[V_{1},H_{0}+V_{0}],V_{1}]+O(\omega^{-4})\label{eq:Heffom}
\end{equation}
This calculation is quite general, which relies only on the specific
time dependence of the perturbation $\cos\omega t\,V_{1}$, but not
on the specific form of $V_{0},V_{1}$ and will be valid also in the
quantum field theory. In the generic case of a single particle, the
effective Hamiltonian is 
\begin{equation}
H_{\mathrm{eff}}=\frac{p_{\phi}^{2}}{2}+V_{0}+\frac{1}{4\omega^{2}}(\partial_{\phi}V_{1})^{2}+O(\omega^{-4})
\end{equation}
Clearly, for the Kapitza pendulum, this yields the same effective
potential $U_{\mathrm{eff}}$ as in (\ref{eq:Ueff}), which is the
quantum analogue of the classical dynamics. In the more general case
when $V_{0}$ and $V_{1}$ are equivalent up to a linear term in the
coordinate, this maps all unstable fix points of $V_{0}$ to stable
ones.

\begin{figure}[h]
\begin{centering}
\includegraphics[width=16cm]{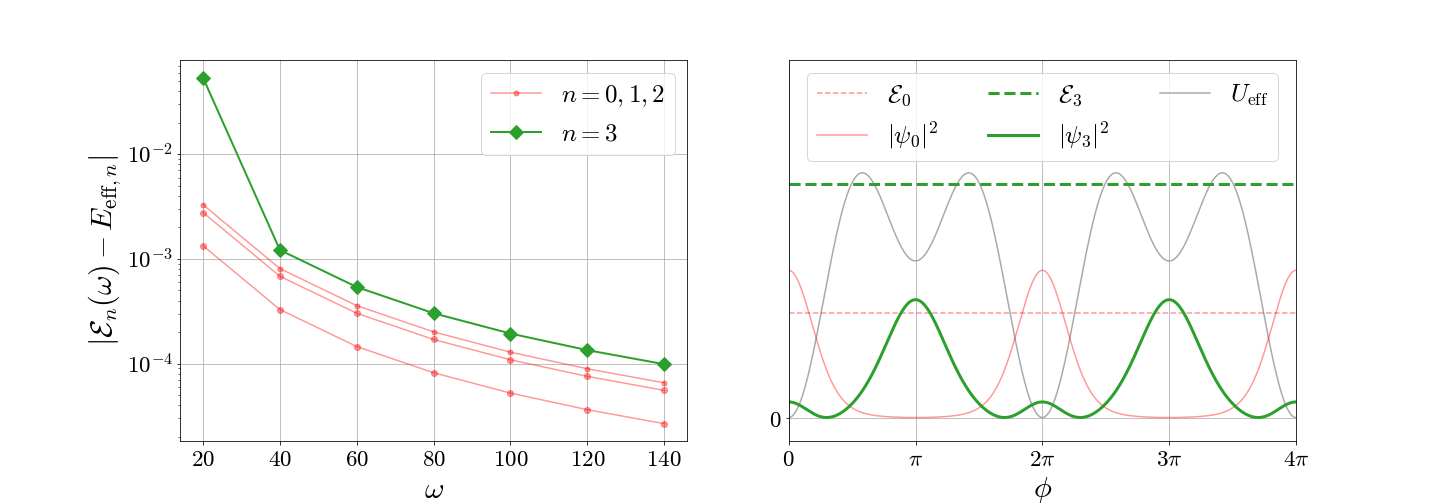}
\par\end{centering}
\caption{Comparison of the large $\omega$ Floquet quasienergy spectrum with
the spectrum of the effective Hamiltonian is shown on the left, while
wave functions localized around the two stable fix points are demonstrated
on the right. Parameters: $c_{1}=\lambda,c_{0}=0.5$, $\lambda=2$.}

\label{fig:QMeff}
\end{figure}

In order to test the above large-frequency approximation, we can determine
numerically the Floquet basis, which satisfies the modified Schrödinger
equation:
\begin{equation}
H_{F}u_{\mathcal{E}}(\phi,t)=\mathcal{E}u_{\mathcal{E}}(\phi,t)\quad;\qquad H_{F}=H-i\partial_{t}
\end{equation}
 Since $u_{\mathcal{E}}(\phi,t)$ is periodic in time, we simply expand
it in Fourier components: $u_{\mathcal{E}}(\phi,t)=\sum_{m}u_{\mathcal{E},m}(\phi)e^{im\omega t}$
such that the eigenvalue problem takes the form 
\begin{equation}
(m\omega+H_{0}+V_{0})u_{\mathcal{E},m}+\frac{1}{2}V_{1}(u_{\mathcal{E},m-1}+u_{\mathcal{E},m+1})=\mathcal{E}u_{\mathcal{E},m}\label{eq:Floquetnum}
\end{equation}
We can further expand $u_{\mathcal{E},m}(\phi)$ in the eigenbasis
of $H_{0}$, and formulate an equation in the double discrete infinite
basis, see Appendix \ref{sec:Large-frequency-expansion} for details.
The numerical solution showed (see Figure \ref{fig:QMeff}) that for
large $\omega$, the effective description is recovered correctly,
and from the eigenvectors, we can recognize states localized around
the upper equilibrium point.

\subsection{The periodically driven sine-Gordon model}

The quantum version of the periodically driven sine-Gordon model shares
the same stucture of the Hamiltonian as the Kapitza pendulum (\ref{eq:Hdriven})
but we are in a quantum field theory where the system is confined
into a finite volume $L$ as
\begin{equation}
H_{0}=\frac{1}{8\pi}\int_{0}^{L}dx:(\partial_{t}\phi)^{2}-(\partial_{x}\phi)^{2}:\quad;\qquad V_{i}=c_{i}\int_{0}^{L}dx:\cos\beta\phi:\label{eq:H0Vi}
\end{equation}
and the normal ordering is defined wrt. the free compactified massless
boson of radius $r=\beta^{-1}$ \cite{Feverati:1998va,Samaj:2013yva}.

\subsubsection{Sine-Gordon model as a perturbed CFT}

\begin{figure}[h]
\begin{centering}
\includegraphics[width=8cm]{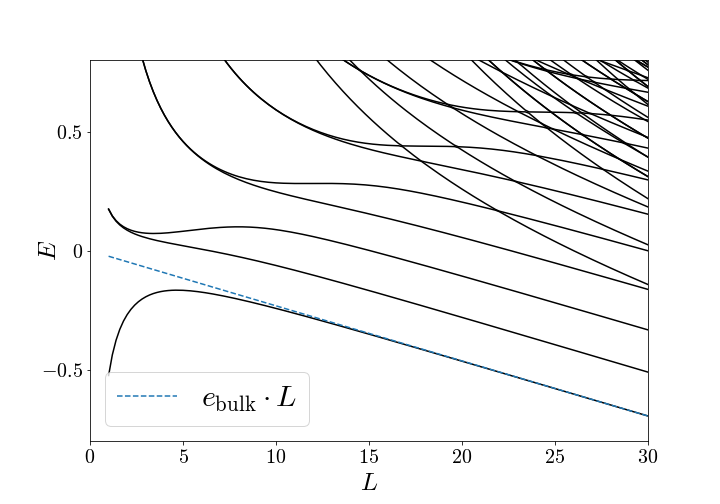}\includegraphics[width=8cm]{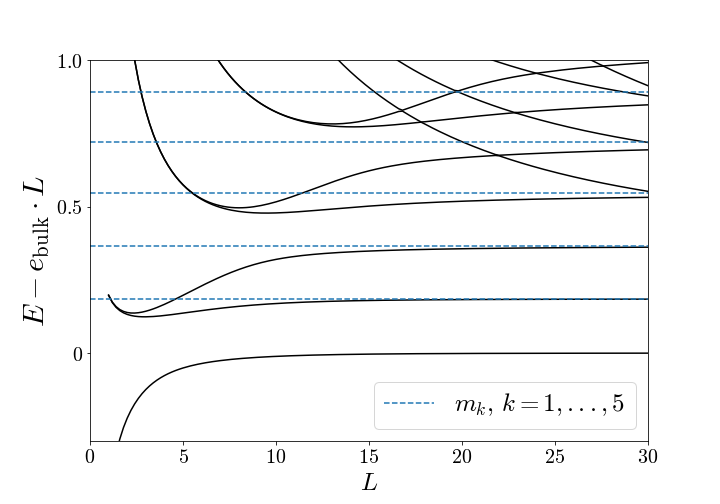}
\par\end{centering}
\caption{Typical TCSA spectrum of the sine-Gordon theory. Here and from now
on we always choose $r=3$, $M=1$. On the left is the raw spectrum,
while on the right we subtract the exactly known bulk groundstate
energy. Breather masses are indicated by dashed lines.}

\label{fig:sGspec}
\end{figure}

Since we work with the free compactified boson, $\phi$ is regarded
as an angle variable with the identification $\phi+2\pi r\equiv\phi$.
This implies that boundary conditions can be labeled by the integer
winding number $m$ as $\phi(x+L,t)=\phi(x,t)+2\pi rm$, which distinguishes
topological sectors closed for time evolutions. The zero mode $\phi_{0}(t)=L^{-1}\int_{0}^{L}\phi(x,t)dx$,
behaves as a single pendulum. In the free theory, its conjugate momentum
$\pi_{0}$, satisfying $[\pi_{0},\phi_{0}]=-i$, has the spectrum
$\pi_{0}\vert n\rangle=\frac{n}{r}\vert n\rangle$ with integer $n$.
The other modes $\phi_{n}$ can be expanded in terms of creation and
annihilation operators, see Appendix \ref{app:TCSA} for details.
Since the free massless boson is a conformal field theory, it is costumary
to work on the plane defined by the conformal mapping $z=e^{i\frac{2\pi}{L}(x+t)}$,
$\bar{z}=e^{-i\frac{2\pi}{L}(x-t)}$. Conformal symmetry implies that
the field separates into independent left and right moving parts and
the full Hilbert space is built over the states with given winding
and momentum numbers by acting with the left and independent right
moving creation operators:
\begin{equation}
\mathcal{H}=\sum_{n,m\in\mathbb{Z}}\mathcal{V}_{n,m}\otimes\bar{\mathcal{V}}_{n,m}\quad;\qquad\mathcal{V}_{n,m}=\{a_{-n_{1}}\dots a_{-n_{k}}\vert n,m\rangle\}\label{eq:Hspace}
\end{equation}
where $[a_{n},a_{m}]=n\delta_{n+m,0}$ and similarly for $\bar{\mathcal{V}}_{n,m}$
with the appropriate changes. The Hamiltonian and the total momentum
are 
\begin{equation}
H_{0}=\frac{2\pi}{L}\left(L_{0}+\bar{L}_{0}-\frac{1}{12}+\pi_{0}^{2}+\frac{m^{2}r^{2}}{4}\right)\quad;\qquad P_{0}=\frac{2\pi}{L}(L_{0}-\bar{L}_{0}+\pi_{0}m)\label{eq:H0P0}
\end{equation}
where $L_{0}=\sum_{k>0}a_{-k}a_{k}$ and $\bar{L}_{0}=\sum_{k>0}\bar{a}_{-k}\bar{a}_{k}$.
The perturbing operator can also be mapped onto the plane. As $:e^{i\beta\phi}:$
is a primary field of weights $(h,\bar{h})=(\frac{\beta^{2}}{2},\frac{\beta^{2}}{2})$
the perturbation takes the form 
\begin{equation}
V_{i}=c_{i}\left(\frac{L}{2\pi}\right)^{1-\beta^{2}}\hat{V}_{1}\quad;\qquad\hat{V}_{1}=\int_{0}^{2\pi}d\theta\frac{1}{2}\left(V^{\beta}(e^{i\theta},e^{-i\theta})+V^{-\beta}(e^{i\theta},e^{-i\theta})\right)\label{eq:V1hat}
\end{equation}
where the normal ordered vertex operator on the plane was introduced
$V^{\beta}(z,\bar{z})=:e^{i\beta\phi(z,\bar{z})}:$. Without the drive,
$c_{1}=0$, normal ordering is enough to regularize the theory for
$\beta^{2}<1$. In this region, the theory contains breather and soliton
excitations. The only dimensionful perturbing parameter $c_{0}$ sets
the scale and it is related to the soliton mass $M$ as \cite{Zamolodchikov:1995xk}:
\begin{equation}
c_{0}=\kappa(h)M^{2-2h}\quad;\qquad\kappa(h)=\frac{2\Gamma(h)}{\pi\Gamma(1-h)}\left(\frac{\sqrt{\pi}\Gamma(\frac{1}{2(1-h)})}{2\Gamma(\frac{h}{2(1-h)})}\right)^{2-2h}
\end{equation}
The truncated conformal space approach can be used the calculate the
spectrum \cite{Yurov:1989yu}. It amounts to truncate the Hilbert
space at a given energy, $E_{\mathrm{cut}},$ and to calculate the
finite matrix representations of the Hamiltonian $H_{0}+V_{0}$ and
then diagonalize them \cite{Feverati:1998va}. A typical spectrum
can be seen on Figure \ref{fig:sGspec}.

Since this is an integrable quantum field theory, the finite size
spectrum is completely known \cite{Destri:1997yz,Feverati:1998dt}.
The bulk energy constant is $e_{\mathrm{bulk}}=-\frac{M^{2}}{4}\tan\frac{\pi p}{2}$,
where $h=\frac{p}{1+p}=\frac{\beta^{2}}{2}$, while the mass of the
$k^{th}$ breather is $m_{k}=2M\sin\frac{\pi pk}{2}$.

For $1<\beta^{2}$ the perturbing operator is not well-defined \cite{Zamolodchikov:1990bk}.
The spectrum does not stabilize in the $E_{\mathrm{cut}}\to\infty$
limit and one has to subtract an $E_{\mathrm{cut}}$-dependent diverging
constant. This can be avoided by analyzing energy differences only.
For higher $\beta$s, even energy differences are not enough to consider
and one has to introduce an $E_{\mathrm{cut}}$-dependent operator
counter term \cite{Zamolodchikov:1990bk,Delfino:1996nf}. Above $\beta^{2}=2$,
the TCSA method cannot be used, as we cross the Kosterlitz-Thouless
phase transition and the perturbation becomes irrelevant.

\subsubsection{Periodic drive}

Let us now analyze the effect of the periodic drive. If the drive
is turned on, we are interested in the Floquet quasienergy spectrum.
The large $\omega$ expansion follows the calculation of the Kapitza
pendulum and leads to the effective Hamiltonian (\ref{eq:Heffom}),
which in dimensionless quantities in our case reads as 
\begin{equation}
H_{\mathrm{eff}}=\frac{2\pi}{L}\left[\hat{H}_{0}+\frac{c_{0}}{2}\left(\frac{L}{2\pi}\right)^{2-\beta^{2}}\hat{V}_{1}+\left(\frac{L}{2\pi}\right)^{2-2\beta^{2}}\frac{c_{1}^{2}}{4\omega^{2}}[[\hat{V}_{1},\hat{H}_{0}],\hat{V}_{1}]\right]+O(\omega^{-4})\label{eq:HeffsG}
\end{equation}
We calculate the commutator $[[\hat{V}_{1},\hat{H}_{0}],\hat{V}_{1}]$
in Appendix \ref{app:TCSA} analytically. As the expression contains
products of operators at the same points, the commutator is not well-defined
and needs to be regularized. By introducing a mode number cutoff,
we keep only the oscillators with mode numbers between $-n_{\mathrm{max}}$
and $n_{\mathrm{max}}$. As a result, the regularized commutator has
the structure
\begin{equation}
[[\hat{V}_{1},\hat{H}_{0}],\hat{V}_{1}]{}_{\mathrm{cut}}\propto an_{\mathrm{max}}^{2\beta^{2}}\mathbb{I}+n_{\mathrm{max}}^{-2\beta^{2}}\hat{V}_{2}+O(n_{\mathrm{max}}^{-2\beta^{2}-1})
\end{equation}
where $\hat{V}_{2}$ corresponds to a perturbation with double frequency:
\begin{equation}
\hat{V}_{2}=\int_{0}^{2\pi}\frac{d\theta}{2}\left(V^{2\beta}(e^{i\theta},e^{-i\theta})+V^{-2\beta}(e^{i\theta},e^{-i\theta})\right)
\end{equation}
The contribution of the identity operator is diverging in the limit
when the cutoff is eliminated ($n_{\mathrm{cut}}\to\infty)$ and should
be renormalized. This can be easily done by considering energy differences
only, and then the large $\omega$ behaviour can be different depending
on how we scale $c_{1}$. We expect the following behaviours:
\begin{itemize}
\item If $c_{1}$ is not scaled with $\omega$, $c_{1}=\lambda$, then the
drive averages out and we should see no effect in the spectrum.
\item If $c_{1}$ is scaled with $\omega$, $c_{1}=\lambda\omega$, then
we have an effective large $\omega$ behaviour, which in energy differences
appears as the spectrum of the two-frequency sine-Gordon model. By
eliminating the regulator $n_{\mathrm{cut}}\to\infty$, the extra
$\cos2\beta\phi$ term in the effective potential scales to zero and
we again should get back the spectrum of the sine-Gordon theory.
\item If, however, we also scale the bare coupling $c_{1}$ with the regulator
as $c_{1}=\lambda\omega E_{\mathrm{cut}}^{\beta^{2}}$, in order to
compensate the factor $n_{\mathrm{max}}^{-2\beta^{2}}$, then we have
a nontrivial large $\omega$ and $E_{\mathrm{cut}}\to\infty$ limit.
In this limit, the volume dependence of the new term in the effective
potential is 
\begin{equation}
\left(\frac{L}{2\pi}\right)^{1-4\beta^{2}}\frac{\lambda^{2}}{4}\left(\frac{E_{\mathrm{cut}}L}{2\pi}\right)^{2\beta^{2}}[[\hat{V}_{1},\hat{H}_{0}],\hat{V}_{1}]
\end{equation}
where the dimensionless cut $\hat{E}_{\mathrm{cut}}=\frac{E_{\mathrm{cut}}L}{2\pi}$
is the analogue of $n_{\mathrm{max}}$ in this scheme and $\hat{E}_{\mathrm{cut}}^{2\beta^{2}}$$[[\hat{V}_{1},\hat{H}_{0}],\hat{V}_{1}]$
has a finite limit corresponding to the double-frequency cosine operator.
The effective theory should then be the two-frequency sine-Gordon
model \cite{Bajnok:2000ar}:
\begin{equation}
H_{\mathrm{eff}}=\frac{2\pi}{L}\left[\hat{H}_{0}+c_{0}\left(\frac{L}{2\pi}\right)^{2-\beta^{2}}\hat{V}_{1}+\left(\frac{L}{2\pi}\right)^{2-4\beta^{2}}c_{2}\hat{V}_{2}\right]
\end{equation}
where $c_{2}\propto\lambda^{2}$ is a scheme-dependent renormalized
coupling.
\end{itemize}
In order to check these behaviours, we develop a novel numerical method.
The idea is to combine TCSA with the numerical approach we used in
the quantum mechanical case. We thus expand in Fourier components
the periodic Floquet wave function in time $u_{\mathcal{E}}(t)=\sum_{m}u_{\mathcal{E},m}e^{im\omega t}$
and solve the Floquet eigenvalue problem (\ref{eq:Floquetnum}) but
keeping in mind that now the operartors $H_{0},V_{0}$ and $V_{1}$
act on the conformal Hilbert space. We use the TCSA method to represent
these operators with finite matrices on the truncated Hilbert space.
Since both the winding number and the momentum is preserved by the
perturbation, we focus on the $m=0$ and $P=0$ sector. The relevant
matrix element of the perturbation are described in Appendix \ref{app:TCSA}.
In the following section we summarize our findings. All physical quantities
are made dimensionless by the soliton mass $M$.

\section{Numerical investigations and results}

In this section, we solve numerically both the time-dependent theory
at high but finite frequency, and its corresponding time-independent
effective theory to support our claims. For a detailed description
of the used methods, see Appendix (\ref{app:TCSA}).

\begin{figure}[h]
\begin{centering}
\includegraphics[width=10cm]{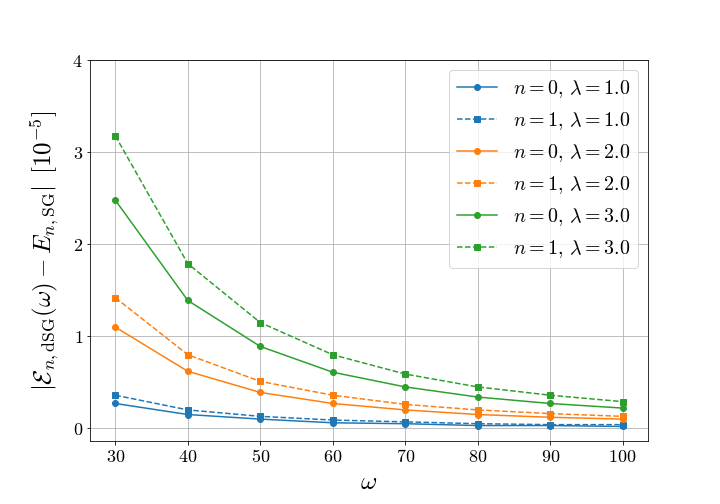}
\par\end{centering}
\caption{The difference between the quasienergies of the driven and the energies
of the original sine-Gordon model for the vacuum and for the first
excited state for various $c_{1}=\lambda$ values as the function
of $\omega$ at volume $ML=14$.}

\label{figure:dsG-sG}
\end{figure}

\subsection{Floquet quasienergy spectrum}

Our first goal is to confirm that the high-frequency expansion is
also valid in the continuum limit. In the case when $c_{1}=\lambda$,
the expansion tells us that the full perturbation vanishes in the
high-frequency limit, therefore, from the quasienergies, we should
obtain the spectrum of the integrable sine-Gordon theory. This is
exactly what we see in Figure \ref{figure:dsG-sG}. Here, we show
that for the vacuum ($n=0$) and the first standing particle ($n=1$),
the difference of the energies tends to zero as the frequency increases
independently of the strength of the perturbation.

\begin{figure}[H]
\begin{centering}
\includegraphics[width=8cm]{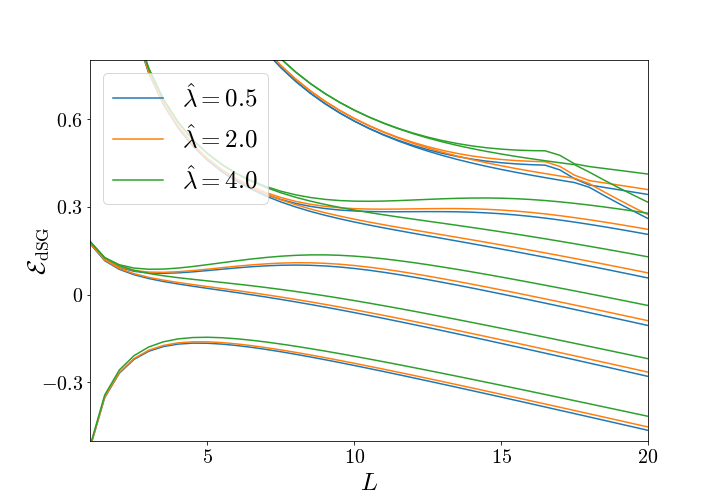}\includegraphics[width=8cm]{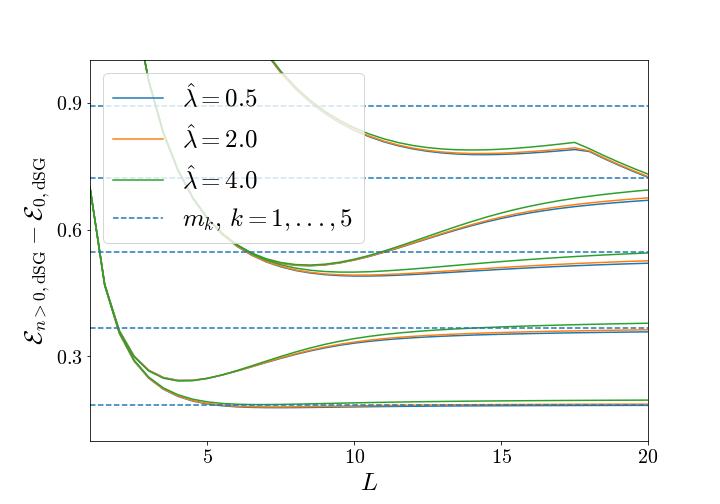}
\par\end{centering}
\caption{Quasienergy spectrum of the driven sine-Gordon theory for various
$\lambda$ drives ($\omega=100$). The raw spectrum is on the left,
while the groundstate energy subtructed is on the right.}

\label{drivensG1}
\end{figure}

Let us now turn to the $c_{1}=\lambda\omega$ case, where we expect
a non-trivial behaviour due to the effect of the $\frac{\lambda^{2}}{4}[[V_{1},H_{0}],V_{1}]$
operator. Looking at the full spectrum on Figure \ref{drivensG1},
we can see that it resembles a meaningful quantum field theory. Knowing
that we are in the $m=0$, $P=0$ sector, we can identify the vacuum,
the standing particles, and the scattering states.

However, comparing it to the sine-Gordon theory, it turns out that
the expected non-trivial contribution is just a constant added to
every energy level, which does not enrich the dynamics. This also
implies that if we subtract the vacuum from every level, we obtain
the unperturbed sine-Gordon theory again, and therefore, we have found
a strong indication that the infinite-frequency effective correction
is a non-vanishing identity in this cutoff scheme too, and the contribution
of the $\hat{V}_{2}$ operator is negligible.

\begin{figure}[h]
\begin{centering}
\includegraphics[width=10cm]{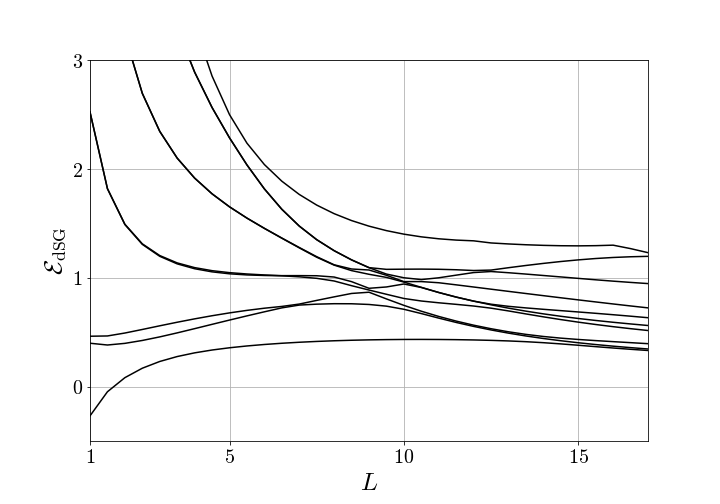}
\par\end{centering}
\caption{Quasienergy levels of the driven sine-Gordon theory with scaled coupling
$c_{1}=\lambda\omega E_{\mathrm{cut}}^{\beta^{2}}$ as the function
of the volume. Parameters: $\hat{\lambda}=14$, $\omega=100$, $\hat{E}_{\mathrm{cut}}=19$}

\label{figure:dsGEscaled}
\end{figure}

In the oscillator mode cutoff scheme, the $c_{1}=\lambda\omega E_{\mathrm{cut}}^{\beta^{2}}$
scaling turned out to be a success in compensating the vanishing $\hat{V}_{2}$
operator, however, this is not trivially achieved in the more physical
energy cutoff scheme. Thus, we now investigate the effect of such
scaling in the time-dependent theory. In Figure \ref{figure:dsGEscaled},
we can see that in this case, a metastable state, a so-called false
vacuum appears, which is an effect known specifically from the two-frequency
sine-Gordon model \cite{Bajnok:2000ar}. The false vacuum is a vacuum-like
energy eigenstate (with linear volume dependence) that has a higher
bulk energy, and in this case, we can interpret it as the analogue
of the stabilized upper fix point of the Kapitza pendulum. This state
can exist in a finite volume., andfor larger and larger volumes, it
approaches a particle line over the real vacuum. Since this theory
is not integrable, the lines avoid each other and the metastable vacuum
decays.

Unfortunately, in the parameter region where the time-dependent theory
shows the stabilization effect the simulations become more resource
demanding and we could not achieve good enough precisions. In order
to overcome this, we compare the time-dependent Floquet spectrum to
that of the effective time-indepent theory.

\begin{figure}[h]
\begin{centering}
\includegraphics[width=10cm]{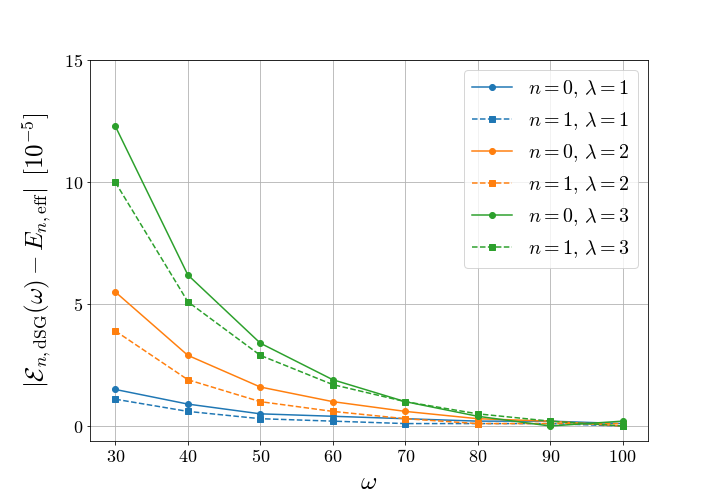}
\par\end{centering}
\caption{The difference between the quasienergies of the driven and energies
of the effective Hamiltonian (\ref{eq:HeffsG}) for the vacuum and
the first excited state for various $c_{1}=\lambda\omega$ values
as the function of $\omega$ at dimensionless volume $ML=14$.}

\label{figure:dsG-sGomega}
\end{figure}

In Figure \ref{figure:dsG-sGomega}, we can see the same behaviour
as we have seen in the quantum mechanical case (Figure \ref{fig:QMeff}):
solving the time-dependent Floquet problem and the time-independent
effective theory yields the same energies towards the high-frequency
limit. We thus in the next subsection focus on a precision analyzis
of the spectrum of the effective Hamiltonian.

\subsection{Effective Hamiltonian}

In this section, we analyze the cases when the drive is scaled with
the frequency $c_{1}=\lambda\omega$ and $\omega$ goes to $\infty$.
The effective Hamiltonian in this limit takes the form 
\begin{equation}
\frac{H_{\mathrm{eff}}}{M}=\frac{2\pi}{ML}\left[\hat{H}_{0}+\frac{\kappa(\beta^{2}/2)}{2}\left(\frac{ML}{2\pi}\right)^{2-\beta^{2}}\hat{V}_{1}+\left(\frac{ML}{2\pi}\right)^{2-2\beta^{2}}\frac{\hat{\lambda}^{2}}{4}[[\hat{V}_{1},\hat{H}_{0}],\hat{V}_{1}]\right]\label{eq:Heffinf}
\end{equation}
where we made the expression dimensionless by dividing by the soliton
mass and $\hat{\lambda}$ is the corresponding dimensionless coupling.
The TCSA method truncates the Hilbert space at a given energy cut
$E_{\mathrm{cut}}$ and represents all operators on this truncated
Hilbert space by finite dimensional matrices. As a consequence, the
commutator $[[\hat{V}_{1},\hat{H}_{0}],\hat{V}_{1}]$ is finite, thus
regularized, but its matrix elements depend on the energy cutoff.
We note that this regularization is not the same which we used in
Appendix (\ref{sec:TCSA}), where we truncated the Hilbert space in
the oscillator types (mode numbers) and not in the energy. In that
case, arbitrarily large energy states could contribute with small
mode numbers.

\begin{figure}[h]
\begin{centering}
\includegraphics[width=14cm]{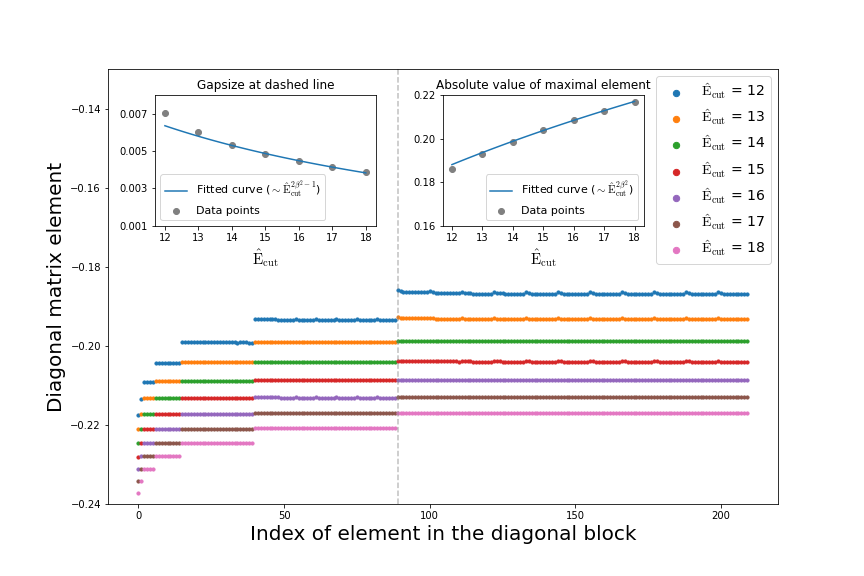}\caption{Diagonal elements of the commutator at different energy cuts, when
projected back to the Hilbert space at $\mathrm{cut}=6$. We see that
at a given $L_{0}$ eigenspace, we basically have an identity operator,
however, the relative cutoff dependences of the different $L_{0}$
eigenspaces are slightly different and are shifted wrt. each other.
The gap between different identity coefficients scales to zero for
larger and larger cuts as shown on the left inset. On the right inset
we can see how the coefficient of the identity scales with the cut.}
\par\end{centering}
\label{diagonal}
\end{figure}

In order to understand the cutoff dependence of the commutator, we
analyze its matrix elements. Technically, we choose a cutoff and construct
the corresponding Hilbert space together with the matrix elements
of $[[\hat{V}_{1},\hat{H}_{0}],\hat{V}_{1}]$. In the next step, we
increase the cutoff as well as the representations of $\hat{H}_{0}$
and $\hat{V}_{1}$ but focus only on the same matrix elements of $[[\hat{V}_{1},\hat{H}_{0}],\hat{V}_{1}]$.
We compare the resulting matrix elements to that of the operators
$\mathbb{I}$ and $\hat{V}_{2}$. The Hilbert space has a tensor product
form composed of the zero mode and the other oscillators. The commutator
in the zero mode space has elements in the diagonal and the second
super/subdiagonal blocks, but none of them depends on the value of
the zero mode. This enables us to investigate the commutator on a
Hilbert space with keeping only the $n=-1,0,1$ sectors. We then analyze
the cutoff dependence in the other oscillators. In our convention,
the cut is an integer which determines the maximal $L_{0}$ eigenvalue.
We first focus on the diagonal block where we expect the appearance
of the identity operator, later we focus on the second subdiagonal
block.

Our results for the diagonal part are presented on Figure \ref{diagonal}.
It seems that the diagonal elements organize themselves into several
distinct identity components, corresponding to the $L_{0}$ eigenvalues.
By increasing the cutoff, the contribution from higher levels decreases
and the gap between these components tends to zero. At the same time,
the cummulated contributions, namely the absolute value of the maximal
element, start to increase. This confirms that in this scheme too,
the effective correction contains an identity operator that diverges
as the cut increases. By using the data in the cut range $14-18$,
we can fit a curve for the coefficient of the identity component in
the form $\sim\hat{E}_{\mathrm{cut}}^{2\beta^{2}}$ with a high precision.
We similarly find that in this range, the gap behaves as $\sim\hat{E}_{\mathrm{cut}}^{2\beta^{2}-1}$.
This confirms that for large cuts the scaling is similar to that of
the mode number cutoff scheme.

\begin{figure}[h]
\begin{centering}
\includegraphics[width=12cm]{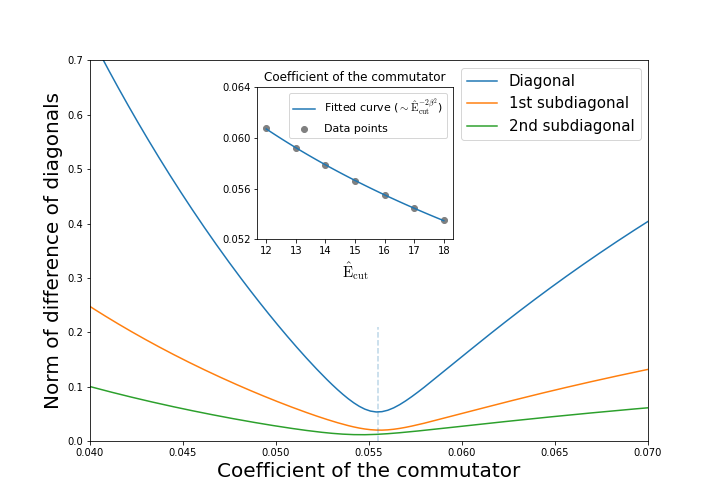}\caption{Comparison of the second subdiagonal block of the commutator with
the analogues term of the double-frequency cosine matrix. We plot
the norm of the diagonal of $x:\cos(2\beta\phi):-[[\hat{V}_{1},\hat{H}_{0}],\hat{V}_{1}]$
as the function of $x$. The minimum in $x$ is basically the same
for the first and second subdiagonals. The inset demostrates the scaling
of the minimum as the function of the cut.}
\par\end{centering}
\label{fig:effvscos2}
\end{figure}

By focusing on the second subdiagonal block, we expect the appearance
of the matrix elements of the double-frequency cosine operator. In
order to confirm this and read off its coefficient, we focus on the
diagonal and the first two subdiagonal elements of the block as the
entries of the cosine matrix decrease strongly from the diagonal.
We take the difference between the diagonals of the commutator and
a multiple of the diagonals of the cosine and calculate the norm of
the difference, which we plot as a function of the multiple factor.
In Figure (\ref{fig:effvscos2}), we can see that there is a specific
coefficient where the diagonals of the matrices coincide with high
precision, which extends also for the first and second subdiagonals.
With the so measured proportionality factor, the remaining subdiagonals
of the commutator all agree at very high precision with the double-frequency
cosine operator. We also measured that this factor decreases with
the cut, meaning that the two-frequency contribution becomes less
and less dominant. Again, we could fit a curve of the form $\sim\hat{E}_{\mathrm{cut}}^{-2\beta^{2}}$
with a very high precision. We therefore conclude that in the energy
cutoff scheme, we find the same operators in the effective Hamiltonian
as in the oscillator cutoff scheme, and their coefficient also behaves
similarly with the increase of the cut.

\begin{figure}[h]
\begin{centering}
\includegraphics[width=10cm]{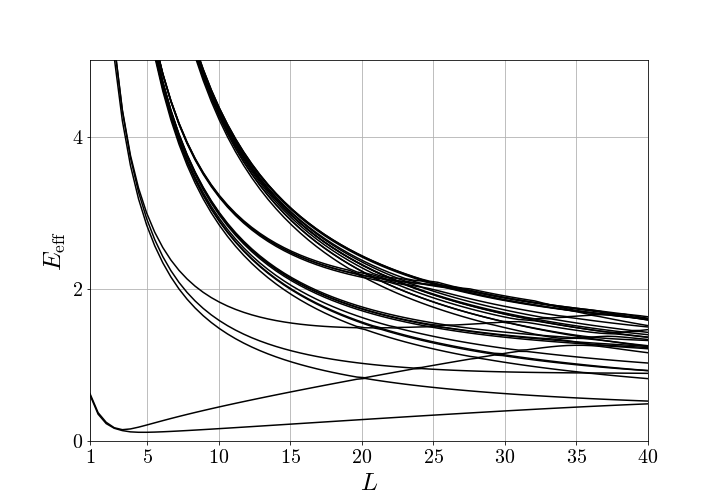}\caption{Spectrum of the effective Hamiltonian with the false vacuum which
confirms that indeed, the upper unstable equilibrium point turned
into an alternate vacuum, just as it happens in the two-frequency
sine-Gordon model. Over this vacuum we can recognize a state with
the same slope, which can be interpreted as a breather-like excitation
over the false vacuum.}
\par\end{centering}
\label{fig:Hefflargecut}
\end{figure}

All these analyzes convinced us that in the case when we scale $\hat{\lambda}$
as $E_{\mathrm{cut}}^{\beta^{2}}$ and consider energy differences
only, we recover the spectrum of the two-frequency model. Indeed using
the operator $H_{\mathrm{eff}}$ with the commutator obtained from
the highest cut available, we obtained the spectrum on Figure \ref{fig:Hefflargecut}.
This shows without any doubts the presence of the stabilized upper
equilibrium point as a false vacuum. We can also recognize an excited
state line parallel with this vacuum, which can be interpreted as
a standing breather-like excitation over the false vacuum. This is
the quantum analogue of the classical solution, which we demonstrated
previously.

\section{Periodically driven perturbed CFTs}

We now extend our analysis from the sine-Gordon model to more general
theories. We still investigate theories with Hamiltonians of the form
of 
\begin{equation}
H=H_{0}+V_{0}+\cos\omega t\,V_{1}
\end{equation}
but the unperturbed part is assumed to be a generic conformal field
theory
\begin{equation}
H_{0}=\frac{2\pi}{L}\hat{H}_{0}\quad;\qquad\hat{H}_{0}=L_{0}+\bar{L}_{0}-\frac{c}{12}
\end{equation}
while the perturbation consists of the integral of relevant spinless
($h_{i}=\bar{h}_{i}$) fields, $\Phi_{0}$ and $\Phi_{1}$: 
\begin{equation}
V_{i}=c_{i}\left(\frac{L}{2\pi}\right)^{1-2h_{i}}\hat{V}_{i}\quad;\qquad\hat{V}_{j}=\int_{0}^{2\pi}\Phi_{j}(e^{i\theta},e^{-i\theta})d\theta
\end{equation}
The effective Hamiltonian in the $\omega\to\infty$ limit takes the
generic form (\ref{eq:Heffinf}), which in dimensionless quantities
reads as 
\begin{equation}
H_{\mathrm{eff}}=\frac{2\pi}{L}\left[\hat{H}_{1}+\left(\frac{L}{2\pi}\right)^{2-4h_{1}}\frac{c_{1}^{2}}{4}[[\hat{V}_{1},\hat{H}_{1}],\hat{V}_{1}]\right]\label{eq:HeffCFT}
\end{equation}
with $\hat{H}_{1}$ being the Hamiltonian of the perturbed CFT without
the drive
\begin{equation}
\hat{H}_{1}=\hat{H}_{0}+c_{0}\left(\frac{L}{2\pi}\right)^{2-2h_{0}}\hat{V}_{0}
\end{equation}
The details of calculating the commutator is relegated to Appendix
(\ref{sec:HeffPCFT}). In the simplest case, when $\Phi_{0}=\Phi_{1}$,
the identity operator always appears in the operator product expansion
(OPE) with a diverging term, which can be renormalized by considering
energy differences only, just as we did in the sine-Gordon case. Assuming
that the OPE starts as 
\begin{equation}
\Phi_{1}(z,\bar{z})\Phi_{1}(0,0)=\frac{1}{(z\bar{z})^{2h_{1}}}+c_{11}^{2}\frac{\Phi_{2}(0,0)}{(z\bar{z})^{2h_{1}-h_{2}}}+\dots
\end{equation}
we can scale the amplitude of the drive as $c_{1}\propto\lambda\omega E_{\mathrm{cut}}^{h_{2}-2h_{1}}$,
in order to compensate the dimensions of $(z\bar{z})^{h_{2}-2h_{1}}$.
The effective large frequency Hamiltonian then takes the form 
\begin{equation}
H_{\mathrm{eff}}=\frac{2\pi}{L}\left[\hat{H}_{0}+c_{0}\left(\frac{L}{2\pi}\right)^{2-2h_{0}}\hat{V}_{0}+c_{2}\left(\frac{L}{2\pi}\right)^{2-2h_{2}}\hat{V}_{2}\right]
\end{equation}
where $c_{2}\propto\lambda^{2}$ is a scheme-dependent effective coupling.
This is the Hamiltonian of a conformal field theory perturbed by two
operators. Thus the periodic drive in the infinite frequency limit
leads to an extra perturbation with an operator, which appears in
the OPE of the driven perturbing operator with itself. These theories
could be systematically analyzed with the methods of \cite{Delfino:1996xp}.
Our result can lead to very interesting phenomena and would open a
new field for the Floquet engineering. In particular, in the critical
Ising theory, it would imply that harmonically changing the magnetic
field would induce an effective temperature perturbation.

\section{Conclusion}

In this paper, we investigated the periodically driven sine-Gordon
quantum field theory, which is considered to be the continuum limit
of coupled many Kapitza pendula. We focused on the large frequency
behaviour and determined the Floquet quasienergy spectrum for various
driving protocols. By exploiting the perturbed CFT nature of the sine-Gordon
model, we combined the TCSA method with the usual numerical Floquet
analysis in order to get a tool to determine the quasienergy spectrum
of the driven sine-Gordon theory for various driving frequencies.
In the large frequency limit, we compared the results with an analytical
calculation based on the large frequency expansion of the effective
Hamiltonian. As we found complete agreement, we analyzed the spectrum
of this effective Hamiltonian, which is non-trivial and $\omega$-independent
if we scale the drive with $\omega$. With this driving protocol,
we observed a uniform cutoff-dependent shift in the energy spectrum
compared to the original sine-Gordon theory. In order to have a non-trivial
effect of the drive in energy differences, we had to scale the drive
also with an appropriate power of the energy cutoff. With this protocol,
the drive had a marked effect on the spectrum, which stabilized when
we increased the cutoff. The resulting quasienergy spectrum agreed
with the spectrum of the two-frequency sine-Gordon theory. In particular,
we observed the signature of another vacuum in the spectrum with higher
bulk energy constant. This other state exists for any volumes, but
does not have an infinite volume limit. It corresponds to the upper
equilibrium point of the coupled pendula, which got dynamically stabilized
by the periodic drive. We even observed an excitation over this false
vacuum.

In our work we could map the large frequency effective behaviour of
the driven system to another equilibrium system with more parameters
and richer dynamics including possible phase transitions. Similar
phenomenon was also analyzed in stochastic driven systems in \cite{Dutta_2003,Dutta_2004}
and our work can be considered its generalization to quantum field
theories. 

Our investigations and methods can be easily generalized to other
periodically driven conformal field theories. Indeed, we already made
the first step into this direction. We calculated the effective Hamiltonian
in the case when the drive is proportional to $\omega$ and an appropriate
power of the energy cutoff. In the case when a conformal field theory
is perturbed with a spinless relevant operator, $\Phi_{1}$ via a
harmonic time-dependent coupling, the resulting theory is time independent
and has two relevant spinless perturbation, $\Phi_{1}$ and $\Phi_{2}$.
The second perturbation $\Phi_{2}$ corresponds to the first non-trivial
operator appearing in the product of $\Phi_{1}$ with itself. In particular,
it implies that by perturbing the critical Ising model with a harmonic
magnetic field, the effective theory contains an additional thermal
perturbation. It would be very interesting to explore the consequences
of our findings in real systems. Also, assuming that we harmonically
drive a conformal field theory, we can have an effective theory with
a single perturbation, which actually can be integrable. One example
is the driven sine-Gordon theory with $c_{0}=0$.

Recently there has been growing interest in periodically driven CFTs
\cite{wen2018floquet,Fan:2019upv,Lapierre:2019rwj,Lapierre:2020ftq,Wen:2020wee}.
In these exactly soluble systems the perturbation is the spatially
modulated energy-momentum density, which is switched on and off periodically.
As a result, the perturbation can be described in terms of the Virasoro
modes and implement conformal transformations, which leave the system
critical. Nevertheless, the stability diagram is extremely rich, which
can be studied via the evolution of the entanglement entropy. In contrast,
in our analysis we focused on the stable large frequency limit of
a relevant perturbation. It would very interesting to use similar
methods, such as the investigation of the entanglement entropy and
map the stability diagram of our phase space. It would be also very
challenging to combine the two type of perturbations. 

In the present paper, we were satisfied by establishing that the appropriately
scaled, harmonically driven sine-Gordon theory is equivalent to the
two-frequency sine-Gordon model. This correspondence, however, can
be explored further. Since the two-frequency model has a plenty of
interesting phenomena including phase transitions and new states in
the spectrum \cite{Delfino:1997ya,Bajnok:2000ar,Takacs:2005fx}, we
expect similar behaviour from the driven model, too. As the driven
sine-Gordon theory can be realized in cold atom experiments, it would
be also very interesting to investigate the experimental consequences
of the appearing two-frequency sine-Gordon model.

\subsection*{Acknowledgements}

We thank Zoltán Rácz for suggesting the problem and the useful discussions
and the NKFIH grant K134946 for support. The work was supported also
by ELKH, while the infrastructure was provided by the Hungarian Academy
of Sciences.

\appendix

\section{Large frequency expansion and a numerical approach \label{sec:Large-frequency-expansion}}

Floquet theorem ensures that the solution of the time-dependent Schrödinger
equation 
\begin{equation}
i\frac{\partial\Psi}{\partial t}=H\Psi\quad;\qquad H=H_{0}+V_{0}+\cos\omega t\,V_{1}
\end{equation}
has the form 
\begin{equation}
\psi_{\mathcal{E}}(\phi,t)=e^{-i\mathcal{E}t}u_{\mathcal{E}}(\phi,t)\quad;\qquad u_{\mathcal{E}}(\phi,t+T)=u_{\mathcal{E}}(\phi,t)
\end{equation}
where the quasienergy $\mathcal{E}$ governs the slow motion in an
effective potential, while the periodic $u_{\mathcal{E}}$ is the
analogue of the fast modulation. In order to find the effective Hamiltonian,
one can make a periodic gauge transformation \cite{PhysRevA.68.013820}
of the form 
\begin{equation}
u_{\mathcal{E}}(\phi,t)=e^{-iF(t)}v_{\mathcal{E}}(\phi)\quad;\qquad H_{\mathrm{eff}}=e^{iF}He^{-iF}-i(\partial_{t}e^{iF})e^{-iF}
\end{equation}
where $F$ and $H_{\mathrm{eff}}$ can be calculated simultaneously
order by order in $\omega^{-1}$: 
\begin{equation}
F=\omega^{-1}F_{1}+\omega^{-2}F_{2}+\dots\quad;\qquad H_{\mathrm{eff}}=H_{\mathrm{eff}}^{(0)}+\omega^{-1}H_{\mathrm{eff}}^{(1)}+\omega^{-2}H_{\mathrm{eff}}^{(2)}+\dots
\end{equation}
by using that 
\begin{equation}
e^{iF}He^{-iF}=H+i[F,H]-\frac{1}{2}[F,[F,H]+\dots\quad;\quad(\partial e^{iF})e^{-iF}=i\partial F-\frac{1}{2}[F,\partial F]+\dots
\end{equation}
and demanding that $H_{\mathrm{eff}}$ is time-independent. At the
leading order, the time dependent part of $H$ can be cancelled by
$F_{1}$ as
\begin{equation}
H_{\mathrm{eff}}^{(0)}=H_{0}+V_{0}\quad;\qquad F_{1}=\sin\omega t\,V_{1}
\end{equation}
At the subleading order, one can ensure $H_{\mathrm{eff}}^{(1)}=0$
by choosing $F_{2}=-i\cos\omega t\,[V_{1},H_{0}+V_{0}]$. At the next
order, $F_{3}$ can compensate terms only with vanishing average leading
to an effective term 
\begin{equation}
H_{\mathrm{eff}}^{(2)}=\frac{1}{4}[[V_{1},H_{0}+V_{0}],V_{1}]
\end{equation}

In order to test the large frequency approximation in the case of
the Kapitza pendulum, we determine numerically the Floquet basis,
which satisfies the modified Schrödinger equation:
\begin{equation}
H_{F}u_{\mathcal{E}}(\phi,t)=\mathcal{E}u_{\mathcal{E}}(\phi,t)\quad;\qquad H_{F}=H-i\partial_{t}
\end{equation}
 Since $u_{\mathcal{E}}(\phi,t)$ is periodic in time, we simply expand
it in Fourier components:
\begin{equation}
u_{\mathcal{E}}(\phi,t)=\sum_{m}u_{\mathcal{E},m}(\phi)e^{im\omega t}
\end{equation}
such that the eigenvalue problem takes the form 
\begin{equation}
(m\omega+H_{0}+V_{0})u_{\mathcal{E},m}+\frac{1}{2}V_{1}(u_{\mathcal{E},m-1}+u_{\mathcal{E},m+1})=\mathcal{E}u_{\mathcal{E},m}
\end{equation}
We can further expand $u_{\mathcal{E},m}(\phi)$ in the eigenbasis
of $H_{0}$: $u_{\mathcal{E},m}(\phi)=\sum_{n}c_{m,n}\vert n\rangle$
which are nothing but the momemtum eigenstates $p_{\phi}\vert n\rangle=n\vert n\rangle$,
which take the form $\vert n\rangle=\frac{1}{\sqrt{2\pi}}e^{in\phi}$.
Since the matrix elements of $V_{0}$ and $V_{1}$ are explicitly
calculable $e^{\pm i\phi}\vert n\rangle=\vert n\pm1\rangle$, the
eigenvalue problem reduces to 
\begin{equation}
(m\omega+\frac{n^{2}}{2})c_{m,n}+\frac{c_{0}}{2}(c_{m,n-1}+c_{m,n+1})+\frac{c_{1}}{4}(c_{m-1,n-1}+c_{m+1,n-1}+c_{m-1,n+1}+c_{m+1,n+1})=\mathcal{E}c_{m,n}
\end{equation}
 Technically, we truncate the tensor product Hilbert space both in
$m$ and $n$ to take values between $\vert m\vert<m_{\mathrm{max}}$
and $\vert n\vert<n_{\mathrm{max}}$ and diagonalize numerically the
finite Hamiltonian. The quasienergy is defined only modulo $\omega$
as we can change $\mathcal{E}\to\mathcal{E}\pm\omega$ by shifting
$m$. This implies that we find any eigenvalue and eigenvector many
times. We thus restrict the eigenspectrum by demanding $0\leq\mathcal{E}<\omega$,
what we call the fundamental region, and order the states wrt. the
time average of the expectation value of the energy $\langle H\rangle$
over one driving period.

\begin{figure}
\begin{centering}
\includegraphics[width=8cm]{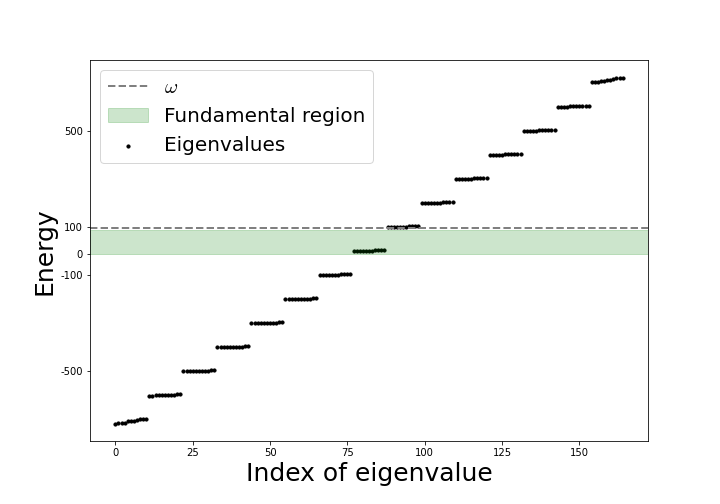}\includegraphics[width=6cm]{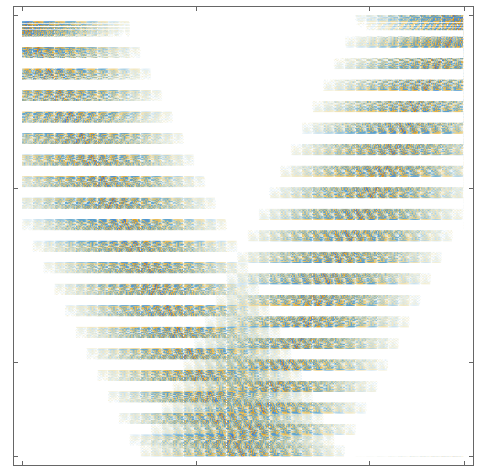}
\par\end{centering}
\caption{On the left, there is a plot of the eigenvalues of the Kapitza pendulum
obtained from our method. The same eigenvalues appear with shifts
of integer multiples of $\omega$, but we only use the central, so-called
fundamental region, which contains all the states with maximal precision.
On the right, there is a plot of the elements of the matrix that diagonalizes
the finite Hamiltonian. Its rows correspond to eigenvectors, and one
can see that practically, different blocks contain the same numbers,
but they belong to different Fourier modes. These shifts eventually
cancel out when one sums up the time-dependent solution, and they
result in the same physical states.}
\end{figure}

\section{TCSA and the driven sine-Gordon model \label{app:TCSA}}

\label{sec:TCSA}

The quantum version of the periodically driven sine Gordon model is
defined by its Hamiltonian (\ref{eq:Floquetnum}) with (\ref{eq:H0Vi}).
Since $\phi$ is an angle variable boundary conditions can be labeled
by the integer winding number $m$ as $\phi(x+L,t)=\phi(x,t)+2\pi rm$,
which distinguishes topological sectors. Within each sector we expand
the field as 
\begin{equation}
\phi(x,t)=2\pi mr\frac{x}{L}+\sum_{n=-\infty}^{\infty}\phi_{n}(t)e^{in\frac{2\pi}{L}x}
\end{equation}
The zero mode $\phi_{0}$ behaves as a single pendulum. In the free
theory its conjugate momentum $\pi_{0}$, satisfying $[\pi_{0},\phi_{0}]=-i$,
has the spectrum $\pi_{0}\vert n\rangle=\frac{n}{r}\vert n\rangle$
with periodic wave functions $e^{i\frac{n}{r}\phi_{0}}$ in $\phi_{0}.$
The other modes $\phi_{n}$ can be expanded in terms of creation and
annihilation operators. Since the free massless boson is a conformal
field theory the field separates into independent left and right moving
parts: 
\begin{equation}
\phi(z,\bar{z})=\varphi(z)+\bar{\varphi}(\bar{z})\quad;\qquad\varphi(z)=\frac{\phi_{0}}{2}-i\pi_{0}\log z+i\sum_{n\neq0}a_{n}\frac{z^{-n}}{n}
\end{equation}
and similarly with $z\leftrightarrow\bar{z}$ and $a\leftrightarrow\bar{a}$
for $\bar{\varphi}$. Canonical commutation relations imply that 
\begin{equation}
[a_{n},a_{m}]=n\delta_{n+m}\quad;\qquad[\bar{a}_{n},\bar{a}_{m}]=n\delta_{n+m}\quad;\qquad[a_{n},\bar{a}_{m}]=0
\end{equation}
and the Hilbert space is built over the states with given winding
and momentum numbers by acting with the left and right moving creation
operators (\ref{eq:Hspace}). The normal ordered Hamiltonian and momentum
can be written in terms of the oscillators as in (\ref{eq:H0P0}).
The perturbing operator, having mapped to the plane, takes the form
(\ref{eq:V1hat}), where the normal ordered vertex operator
\begin{equation}
V^{\beta}(z,\bar{z})=:e^{i\beta\phi(z,\bar{z})}:=V_{0}^{\beta}V_{-}^{\beta}V_{+}^{\beta}
\end{equation}
can be factorized into the zero mode 
\begin{equation}
V_{0}^{\beta}(z,\bar{z})=e^{i\beta\phi_{0}}(z\bar{z})^{\beta\pi_{0}}
\end{equation}
and the creation-annihilation parts 
\begin{align}
V_{\mp}^{\beta}(z,\bar{z})\quad & =\prod_{n>0}e^{\pm\beta a_{\mp n}\frac{z^{\pm n}}{n}}\prod_{n>0}e^{\pm\beta\bar{a}_{\mp n}\frac{\bar{z}^{\pm n}}{n}}
\end{align}

In calculating the effective Hamiltonian (\ref{eq:HeffsG}) we need
to evaluate $[[\hat{V}_{1},\hat{H}_{0}],\hat{V}_{1}]$, where the
dimension-less conformal Hamiltonian is $H_{0}=\frac{2\pi}{L}\hat{H}_{0}$
and the dimensionless part of $V_{1}$ is 
\begin{equation}
\hat{V}_{1}=\int_{0}^{2\pi}\frac{d\theta}{2}\left(V^{\beta}(e^{i\theta},e^{-i\theta})+V^{-\beta}(e^{i\theta},e^{-i\theta})\right)
\end{equation}
We start to calculate $[\hat{H}_{0},V^{\beta}(z,\bar{z})]$, with
$V^{\beta}(z,\bar{z})=V_{0}^{\beta}V_{-}^{\beta}V_{+}^{\beta}$. Using
the commutation relations we obtain
\begin{align}
[\hat{H}_{0},V^{\beta}(z,\bar{z})] & =[\pi_{0}^{2},V_{0}^{\beta}]V_{-}^{\beta}V_{+}^{\beta}+\sum_{n>0}(V_{0}^{\beta}[a_{-n}a_{n},V_{-}^{\beta}]V_{+}^{\beta}+V_{0}^{\beta}V_{-}^{\beta}[a_{-n}a_{n},V_{+}^{\beta}])\nonumber \\
 & =\beta(J_{-}(z,\bar{z})V^{\beta}(z,\bar{z})+V^{\beta}(z,\bar{z})J_{+}(z,\bar{z}))\nonumber \\
 & =\beta^{2}V^{\beta}(z,\bar{z})+(z\partial+\bar{z}\bar{\partial})V^{\beta}(z,\bar{z})
\end{align}
where we introduced 
\begin{equation}
J_{\mp}(z,\bar{z})=\pi_{0}+\sum_{n>0}(a_{\mp n}z^{\pm n}+\bar{a}_{\mp n}\bar{z}^{\pm n})
\end{equation}
In calculating the next commutator we obtain
\begin{equation}
[J_{\mp}(z,\bar{z}),V^{\beta}(w,\bar{w})]=\beta\left(1+\sum_{n>0}z^{\pm n}w^{\mp n}+\sum_{n>0}\bar{z}^{\pm n}\bar{w}^{\mp n}\right)V^{\beta}(w,\bar{w})
\end{equation}
 In particular, when we calculate the effective Hamiltonian we need
terms of the form 
\begin{align}
[[\hat{H}_{0},V^{\alpha}(z,\bar{z})],V^{\beta}(w,\bar{w})] & =\alpha[J_{-}(z,\bar{z})V^{\alpha}(z,\bar{z})+V^{\alpha}(z,\bar{z})J_{+}(z,\bar{z}),V^{\beta}(w,\bar{w})]\nonumber \\
 & ={\rm terms\,\,multiplying\,\,}[V^{\alpha}(z,\bar{z}),V^{\beta}(w,\bar{w})]\nonumber \\
 & \,\,\,\,+\alpha\beta\bigl(\sum_{n}z^{n}w^{-n}+\sum_{n}\bar{z}^{n}\bar{w}^{-n}\bigr)V^{\alpha}(z,\bar{z})V^{\beta}(w,\bar{w})
\end{align}
Let us note that $V^{\alpha}(z,\bar{z})=:e^{i\alpha\phi(z,\bar{z})}:\,\propto e^{i\alpha\phi(z,\bar{z})}$
where the proportionality is a regulator dependent c-number. Since
the equal time commutator of $\phi$ with itself is vanishing 
\begin{equation}
[\phi(x,t),\phi(x',t)]=0
\end{equation}
the commutator of two vertex operator does not contribute. Furthermore
we know the operator product expansion of the vertex operator
\begin{equation}
V^{\alpha}(z,\bar{z})V^{\beta}(w,\bar{w})=(z-w)^{\alpha\beta}(\bar{z}-\bar{w})^{\alpha\beta}\left(V_{\alpha+\beta}(w,\bar{w})+(z-w)\partial V_{\alpha+\beta}(w,\bar{w})+\dots\right)
\end{equation}
where we indicated the first descendant. Recalling that we have to
integrate eventually with $z=e^{i\theta}$, $\bar{z}=e^{-i\theta}$
and $w=e^{i\theta'}$ and $\bar{w}=e^{-i\theta'}$ on the unit circle:
\begin{equation}
\int_{0}^{2\pi}\int_{0}^{2\pi}d\theta d\theta'\sum_{n}e^{in(\theta-\theta')}(2-e^{i(\theta-\theta')}-e^{-i(\theta-\theta')})^{\alpha\beta}V_{\alpha+\beta}(e^{i\theta'},e^{-i\theta'})+\dots\label{eq:sumeint}
\end{equation}
If we would sum up in $n$ from $-\infty$ to $\infty$ , then we
could use that $\sum_{n}e^{in\theta}=2\pi\delta(\theta)$. This actually
implies that the integral is $0$ if $\alpha\beta>0$ or it is $\infty$
if $\alpha\beta<0$. In our case $\alpha=\pm\beta$ and we always
face a divergent behaviour. In order to regularize this we introduce
a cutoff in $n$ and analyze the dependence on this cut. The $\theta$
integral is periodic and we can evaluate as 
\begin{equation}
\int_{0}^{2\pi}d\theta e^{in\theta}(2-2\cos\theta)^{\alpha\beta}=\frac{\pi\sec(\pi\alpha\beta)\Gamma(n-\alpha\beta)}{\Gamma(-2\alpha\beta)\Gamma(\alpha\beta+n+1)}
\end{equation}
It is symmetric in $n$ and the large $n$ behaviour is 
\begin{equation}
\frac{\Gamma(n-\alpha\beta)}{\Gamma(\alpha\beta+n+1)}=n^{-1-2\alpha\beta}\left(1-\frac{1+2\alpha\beta}{n}+\dots\right)
\end{equation}
This implies that the sum behaves as 
\begin{equation}
\sum_{n=-n_{\mathrm{max}}}^{n_{\mathrm{max}}}\frac{\Gamma(n-\alpha\beta)}{\Gamma(\alpha\beta+n+1)}\propto n_{\mathrm{max}}^{-2\alpha\beta}
\end{equation}
We need to analyze two cases. For $\alpha=\beta$ the sum is convergent
and goes to zero. For the leading term in the OPE it goes to zero
quite slowly as $\beta^{2}<1$, however, for descendant the exponent
is shifted by an integer, thus it goes to zero much faster. For $\alpha=-\beta$
the sum is divergent. For the leading term in the OPE, which contains
the identity, it diverges again slowly as $\beta^{2}<1$. For descendants
it is again convergent.

Let us point out that the introduction of the cutoff $n_{\mathrm{max}}$
physically means that we keep only oscillators, which create particles
with conformal energies smaller than $n_{\mathrm{max}}.$ We allow
however an arbitrary number of these particles showing that the cut
is not an energy cut. In this regularization scheme the effective
Hamiltonian has the structure 
\begin{equation}
H_{\mathrm{eff}}^{(2)}\propto\frac{1}{4}[[\hat{V}_{1},\hat{H}_{0}],\hat{V}_{1}]\propto an_{\mathrm{max}}^{2\beta^{2}}\mathbb{I}+n_{\mathrm{max}}^{-2\beta^{2}}\hat{V}_{2}+O(n_{\mathrm{max}}^{-2\beta^{2}-1})
\end{equation}
where $\hat{V}_{2}$ corresponds to a perturbation with double frequency:
\begin{equation}
\hat{V}_{2}=\int_{0}^{2\pi}d\theta\frac{1}{2}\left(V^{2\beta}(e^{i\theta},e^{-i\theta})+V^{-2\beta}(e^{i\theta},e^{-i\theta})\right)
\end{equation}

Finally, let us develop a numerical method to check these behaviours.
The idea is the same as in the quantum mechanical case. We Fourier
expand the periodic Floquet wave function in time $u_{\mathcal{E}}(t)=\sum_{m}u_{\mathcal{E},m}e^{im\omega t}$
and solve the Floquet eigenvalue problem (\ref{eq:Floquetnum}) but
now the operators $H_{0},V_{0}$ and $V_{1}$ act on the conformal
Hilbert space. We use the TCSA method to represent these operators
with finite matrices on the truncated Hilbert space. The relevant
matrix element of the perturbation are as follows. The zero mode has
matrix elements
\begin{equation}
\langle n,m\vert V_{0}^{\pm\beta}\vert n',m'\rangle=\delta_{m,m'}\delta_{n,n'\pm1}
\end{equation}
while for the $s^{th}$ oscillator we have 
\begin{equation}
\langle n,m\vert a_{s}^{l}e^{\pm\beta a_{-s}/s}e^{\mp\beta a_{s}/s}a_{-s}^{k}\vert n,m\rangle=\sum_{i=\mathrm{max}(0,k-l)}^{k}{k \choose i}(-1)^{i}s^{k-i}(k-i)!(\pm\beta)^{l-k+2i}{l \choose k-i}
\end{equation}
These states should be normalized with the square root of $\langle0\vert a_{s}^{l}a_{-s}^{k}\vert0\rangle=\delta_{l,k}k!s^{k}$.
The truncated Hilbert space is defined to keep states below a given
energy cutoff. Thus this is not equivalent to keep oscillators below
a given $n_{\mathrm{max}}$ as even a single oscillator has states
with arbitrarily large energies.

Although we were only insterested in the infinite frequency limit,
where it is enough to diagonalize the much smaller matrices of the
effective theory, our method is applicable to general frequencies,
where the dynamics can be much more complex. In this case, the time-dependence
increases the size of the matrices such that implementing an algorithm
that can handle them can easily become a challenging task.

In our implementation, we used the PRIMME \cite{PRIMME,svds_software}
C-package that was specifically designed for large-scale eigenvalue
problems. PRIMME has an interface that only needs a function that
realizes a matrix-vector multiplication, thus giving the user the
complete freedom of representing the matrix. This allows us to exploit
the double tensor product structure of the Hilbert space so that we
only have to store a single block of elements of $V_{0}$, which reduces
the memory costs to a level that is also sufficient for a general
purpose computer.

\section{Calculation of the effective Hamiltonian for generic pertrubed CFTs}

\label{sec:HeffPCFT}

In this appendix we calculate the effective Hamiltonian for generic
perturbation of CFTs (\ref{eq:HeffCFT}). In doing so we recall that
\begin{equation}
[L_{0},\Phi(z,\bar{z})]=h\Phi(z,\bar{z})+z\partial_{z}\Phi(z,\bar{z})
\end{equation}
and similar relations for $\bar{L}_{0}$ with $\bar{z}$, which are
considered to be independent, but put to $z=e^{i\theta}$ and $\bar{z}=e^{-i\theta}$
at the end of the calculation. The standard trick in calculating the
equal time commutator \cite{Ginsparg:1988ui} is to exploit the fact
that products of operators are well-defined only when they are radially
ordered, the analogue of time ordering after the exponential mapping.
Thus the commutator can be replaced by radially ordered products,
i.e. by deforming the $z_{1}$ integration around the unit circle
$C_{1}$: a bit increasing the radius in the first and decreasing
in the second term as:
\begin{equation}
\oint_{C_{1}}dz_{1}\oint_{C_{1}}dz_{2}[\Phi_{1}(z_{1},\bar{z}_{2}),\Phi_{2}(z_{2},\bar{z}_{2})]=\left(\oint_{C_{1+\epsilon}}\oint_{C_{1}}-\oint_{C_{1-\epsilon}}\oint_{C_{1}}\right)dz_{1}dz_{2}\,R(\Phi_{1}(z_{1},\bar{z}_{1}),\Phi_{2}(z_{2},\bar{z}_{2}))
\end{equation}
As usual, we do not indicate radial ordering explicitly any more.
Radially ordered products have the operator product expansion (OPE)
\begin{equation}
\Phi_{1}(z_{1},\bar{z}_{1})\Phi_{2}(z_{2},\bar{z}_{2})=c_{12}^{i}\frac{1}{((z_{1}-z_{2})(\bar{z}_{1}-\bar{z}_{2}))^{h_{1}+h_{2}-h_{i}}}\Phi_{i}(z_{2},\bar{z}_{2})+\dots
\end{equation}
Since $z$ and $\bar{z}$ are considered to be independent, not the
complex conjugate of each other, when we take $\vert z_{1}\vert>1$
we have to also take $\vert\bar{z}_{1}\vert>1$. This implies that
the contributions from $z_{1}=(1+\epsilon)e^{i\theta_{1}},\bar{z}_{1}=(1+\epsilon)e^{-i\theta_{1}}$
and from $z_{1}=(1-\epsilon)e^{i\theta_{1}},\bar{z}_{1}=(1-\epsilon)e^{-i\theta_{1}}$
cancel each other. This is not true when we perform the same calculations
with $z_{1}\partial_{z_{1}}\Phi_{1}$ (or with $\bar{z}_{1}\partial_{\bar{z}_{1}}\Phi_{1}$)
instead of $\Phi_{1}$ as these derivatives introduce an unbalanced
pole of the form of $1/(z_{1}-z_{2})$ or $1/(\bar{z}_{1}-\bar{z}_{2})$).
The difference of the integration contours results in a delta function
$2\pi i\delta(z_{1}-z_{2})$. This is the analogue of the $\delta(\theta_{1}-\theta_{2})=\sum_{n}e^{in(\theta_{1}-\theta_{2})}$
term in (\ref{eq:sumeint}). When calculated without an energy cutoff,
the delta function puts $\theta_{1}$ to $\theta_{2}$ leading to
zero/infinite depending on the sign of $h_{1}+h_{2}-h_{i}$ in complete
analogy with the similar results in the sine-Gordon theory. In order
to get a finite and non-trivial result, we have to scale the amplitude
of the drive $c_{1}$ appropriately with the cutoff.

Depending on the rich spectrum of the CFT we can expect numerous interesting
cases. In the following, however, we focus only on the simplest, namely
when $\Phi_{0}=\Phi_{1}$. In this case the identity operator always
appears in the OPE with a diverging term, which can be renormalized
by considering energy differences only, just as we did in the sine-Gordon
case. Assuming that the OPE starts as 
\begin{equation}
\Phi_{1}(z,\bar{z})\Phi_{1}(0,0)=\frac{1}{(z\bar{z})^{2h_{1}}}+c_{11}^{2}\frac{\Phi_{2}(0,0)}{(z\bar{z})^{2h_{1}-h_{2}}}+\dots
\end{equation}
we can scale the amplitude of the drive as $c_{1}\propto\lambda\omega E_{\mathrm{cut}}^{h_{2}-2h_{1}}$,
in order to compensate the dimensions of $(z\bar{z})^{h_{2}-2h_{1}}$.
The effective large frequency Hamiltonian then takes the form 
\begin{equation}
H_{\mathrm{eff}}=\frac{2\pi}{L}\left[\hat{H}_{0}+c_{0}\left(\frac{L}{2\pi}\right)^{2-2h_{0}}\hat{V}_{0}+c_{2}\left(\frac{L}{2\pi}\right)^{2-2h_{2}}\hat{V}_{2}\right]
\end{equation}
where $c_{2}\propto\lambda^{2}$ is a scheme-dependent effective coupling.

\bibliographystyle{elsarticle-num}
\bibliography{paper}

\end{document}